\documentclass[aps,prd,showpacs,notitlepage,nofootinbib,superscriptaddress,floatfix,showkeys,twocolumn]{revtex4-2}

\usepackage[parfill]{parskip}    
\usepackage{graphicx}
\usepackage{subfigure}
\usepackage{amssymb}
\usepackage{bm}
\usepackage{epstopdf} 
\usepackage{amsmath}
\usepackage{mathrsfs}
\usepackage{varioref}
\usepackage{mathtools}
\usepackage{hyperref} 
\hypersetup{
	bookmarks=false,         
	pdfstartview={FitH},    
	colorlinks=true,       
	linkcolor=cyan,          
	citecolor=green,        
	filecolor=blue,      
	urlcolor=blue           
}

\DeclareFontFamily{OT1}{pzc}{}
\DeclareFontShape{OT1}{pzc}{m}{it}%
{<-> s * [0.900] pzcmi7t}{}
\DeclareMathAlphabet{\mathscr}{OT1}{pzc}%
{m}{it}
\labelformat{section}{Section #1} 
\labelformat{subsection}{Section #1} 
\labelformat{equation}{Eq.~(#1)} 
\labelformat{figure}{Fig.~#1} 
\labelformat{subfigure}{Fig.~\thefigure#1} 
\labelformat{table}{Tab.~#1} 
\begin{document}
\title{The validity of separate-universe approach in transient ultra-slow-roll inflation}
 \author{Rathul Nath Raveendran}
 \email{rathulnath.r@gmail.com}
 \affiliation{Department of Physics, Indian Institute of Science,
C. V. Raman Road, Bangalore 560012, India}

\begin{abstract}
We investigate the breakdown of the separate-universe approximation during transitions in transient ultra-slow-roll inflation by analyzing the evolution of the comoving curvature perturbation ${\cal R}$ and its conjugate momentum $\Pi$. It is well known that spatial gradient terms lead to a failure of this approximation, particularly at the transition from slow-roll to ultra-slow-roll phase. We show that a similar breakdown also occurs during the second transition back to slow-roll when considering the evolution of $\Pi$. Interestingly, while the homogeneous solution for $\Pi$ accurately captures the dynamics across the first transition, it is the homogeneous solution for ${\cal R}$ that becomes valid across the second. Furthermore, we demonstrate that the spatial curvature term introduced in the extended $\delta N$ formalism of \cite{Artigas:2024ajh} can be interpreted as arising from the contribution of $\Pi$ to the energy density perturbation. Importantly, this modification of the local Hubble parameter is valid only when the first slow-roll parameter is both small and strictly constant.

\end{abstract}
\maketitle
\flushbottom
\section{Introduction}

According to the standard theory of slow-roll inflation, inhomogeneities in the universe originate from quantum fluctuations~\cite{Guth:1982ec,Starobinsky:1982ee,Sasaki:1986hm,Mukhanov:1988jd,Martin:2012ua}. As the universe expands, the physical wavelengths associated with these fluctuations grow and eventually exceed the Hubble radius of the background universe. Once this happens, the inhomogeneous universe can be effectively described as a collection of locally homogeneous patches—each evolving as a separate-universe~\cite{Sasaki:1995aw,Wands:2000dp,Rigopoulos:2003ak,Tanaka:2021dww}. This concept forms the foundation of the $\delta N$ formalism, which has been extensively used to study the long-wavelength evolution of cosmological perturbations~\cite{Salopek:1990jq,Sasaki:1998ug,Lyth:2005fi,Abolhasani:2013zya,Chen:2013eea,Cruces:2022dom,Artigas:2021zdk, Artigas:2025nbm}. 

Recent studies have shown that the separate-universe approach transiently breaks down during transitions between slow-roll and ultra-slow-roll phases in certain inflationary models, particularly due to the significant role played by spatial gradient terms in the evolution of the comoving curvature perturbation\cite{Leach:2001zf,Naruko:2012fe,Jackson:2023obv,Domenech:2023dxx}. This failure limits the applicability of the standard $\delta N$ formalism, which assumes locally homogeneous evolution. To address this issue, a modified framework has been proposed wherein each Hubble patch is modeled as a local Friedmann–Lemaître–Robertson–Walker (FLRW) universe with an effective spatial curvature term~\cite{Artigas:2024ajh}. The resulting extended $\delta N$ formalism accurately describes the superhorizon evolution of curvature perturbations during transient ultra-slow-roll inflation, thereby restoring the predictive power of the separate-universe approach in these regimes.

In this work, we analyze the breakdown of the separate-universe approach and assess the validity of extended $\delta N$ formalism during transitions in transient ultra-slow-roll inflation by tracing the evolution of the comoving curvature perturbation ${\cal R}$ and its conjugate momentum $\Pi$ (see Refs.~\cite{Martin:2012pe,Choudhury:2013woa,Dimopoulos:2017ged,Bhaumik:2019tvl,Ragavendra:2020sop,Raveendran:2022dtb,Choudhury:2023hfm,Raveendran:2023auh} for more details on ultra-slow-roll inflation). In~\ref{sec:linear-perturbation}, we set up the essential equations of linear perturbations relevant to our analysis.~\ref{sec:background perturbations} discusses the perturbations of key background quantities and their connection to the homogeneous solutions of linear perturbations. In~\ref{sec:USR}, we investigate the separate-universe approach and the $\delta $N formalism within a specific model of transient ultra-slow-roll inflation. Finally, in~\ref{sec:summary}, we summarize our findings and outline their implications.

Conventions and Notations: We adopt natural units by setting $\hbar = c = 1$, and define the reduced Planck mass as $M_{\rm Pl} = (8\pi G)^{-1/2}$. Following standard conventions, derivatives with respect to cosmic time $t$ are denoted by a dot while derivatives with respect to conformal time $\eta$ are indicated by a prime.

\section{Evolution of perturbations} \label{sec:linear-perturbation}

To analyze scalar perturbations in an expanding universe, we begin by considering fluctuations around the homogeneous and isotropic FLRW background metric. If we take into account the scalar perturbations to the background 
metric, then the FLRW line-element, in general,  
can be written as~\cite{Kodama:1984ziu,MUKHANOV1992203}
\begin{eqnarray}
{\rm d} s^2
&=& -\left(1+2\, A\right)\,{\rm d} t ^2 
+ 2\, a(t)\, (\partial_{i} B)\, {\rm d} t\, {\rm d} x^i \\ \nonumber
&&+a^{2}(t)\, \left[(1-2\, \psi)\, \delta _{ij}
+ 2\, \left(\partial_{i}\, \partial_{j}E \right)\right]\,
{\rm d} x^i\, {\rm d} x^j, \label{eq:metric}
\end{eqnarray}
where $A$, $B$, $\psi$ and $E$ represent the four degrees of freedom associated with scalar perturbations in the metric, which depend on time as well as space.

At linear order in the perturbations, the Einstein field equations can be linearized and yield the following set of coupled differential equations for the metric potentials and energy-momentum perturbations~\cite{Mukhanov:1990me,Sriramkumar:2009kg}:
\begin{subequations}\label{eq:fo-ee}
\begin{eqnarray}
	3\,H\,\left(H\,A + \dot{\psi}\right) 
	- \frac{
	\nabla^2}{a^2}\left[\psi - H\,\sigma\right]
	&=& -\frac{\delta \rho}{2\,M_{\rm Pl}^2},\\
	\partial_i\left(H\,A + \dot{\psi}\right) 
	&=& -\frac{\delta T^0_i}{2\,M_{\rm Pl}^2},\\
	A - \psi + \frac{1}{a}\,\left(a\,\sigma\right)^{\cdot}&=&0\label{eq:fo-ee-3}.
\end{eqnarray}
\end{subequations}
The \ref{eq:fo-ee-3} arises from the assumption that the stress-energy tensor contains no anisotropic stress at linear order, a condition often satisfied for scalar field matter sources.
The components of energy-momentum tensor, $\delta T^\mu_\nu$ for a scalar field can be calculated to be
\begin{subequations}
\begin{eqnarray}\label{eq:delta-T}
\frac{\delta \rho}{2\,M_{\rm Pl}^2} &=& \epsilon_1\,H^2 \left[
\left(\frac{\delta \phi}{\dot{\phi}}\right)^{\cdot} - A 
-3\,H \left(\frac{\delta \phi}{\dot{\phi}}\right)\right],\quad\\
\frac{\delta p}{2\,M_{\rm Pl}^2} &=& \epsilon_1\,H^2 \left[
\left(\frac{\delta \phi}{\dot{\phi}}\right)^{\cdot} - A 
-3\,H \frac{\dot{p}}{\dot{\rho}}\left(\frac{\delta \phi}{\dot{\phi}}\right)\right],\quad\\
\frac{\delta T^0_i}{2\,M_{\rm Pl}^2}&=& -\partial_i \left(\epsilon_1\,H^2\,\frac{\delta \phi}{\dot{\phi}}\right).
\end{eqnarray}
\end{subequations}

To simplify the description of cosmological perturbations and isolate the physically relevant degrees of freedom, it is useful to define following three gauge invariant quantities as~\cite{Bardeen:1980kt,Kodama:1984ziu}
\begin{subequations}
\begin{eqnarray}
{\cal R} &=& \psi + H \,\frac{\delta \phi}{\dot{\phi}},\label{eq:R-def}\\
\Sigma &=& \sigma + \frac{\delta \phi}{\dot{\phi}},\label{eq:Sigma-def}\\
{\cal A} &=& A - \left(\frac{\delta \phi}{\dot{\phi}}\right)^{\cdot}.\label{eq:A-def}
\end{eqnarray}  
\end{subequations}

In the slow-roll approximation, inflationary dynamics are often described using the Hubble slow-roll parameters. These are defined recursively as:
\begin{equation}\label{eq:def-epsilons}
\epsilon_1 = - \frac{{\rm d\, log}H}{{\rm d} N}\, ,\quad
\epsilon_{n+1} = \frac{{\rm d\, log}\epsilon_n}{{\rm d} N} \,,
\end{equation}
where $N$ denotes the number of e-folds. Additionally, we define the quantity $z \equiv a\, \sqrt{2\, \epsilon_1}\,M_{\rm Pl}$ which frequently appears in the second-order action for scalar perturbations. The variable $\Pi \equiv z^2\, {\cal R}'$ is defined as the conjugate momentum associated with the comoving curvature perturbation ${\cal R}$~\cite{Brustein:1998kq,Raveendran:2023auh}.

Using these definitions, the Einstein equations in gauge-invariant form become:
\begin{subequations}\label{eq:Es-r-a-sigma}
\begin{eqnarray}
{\cal R}' - \frac{\nabla^2}{a \, H\,\epsilon_1} \, \left( {\cal R} - H \, \Sigma\right)&=&0,\label{eq:Es-r}\\
{\cal R}' + a\, H\, {\cal A} &=&0,\label{eq:Es-a}\\
\Sigma' +a\,  H\, \Sigma - \frac{\nabla^2}{a \, H\,\epsilon_1} \, \left( {\cal R} - H \, \Sigma\right)-a \, {\cal R}&=&0.\label{eq:Es-sigma}
\end{eqnarray}
\end{subequations}
The conjugate momentum, $\Pi$ now be expressed directly in terms of ${\cal A}$ as
\begin{equation}\label{eq:Pi-in-sA}
    \Pi=- z^2 a\,H\, {\cal A} .
\end{equation}
Moreover, the perturbations to energy density and pressure can be reformulated in terms of $\Pi$ and the scalar field fluctuation as:
\begin{subequations}
\begin{eqnarray}
\label{eq:drho-in-Pi-dphi}
    \delta \rho &=& \frac{H}{a^3} \Pi - 6\, M_{\rm Pl}\,H^2\, \epsilon_1\,\left(\frac{\delta \phi}{\sqrt{2 \, \epsilon_1}}\right), \\
 \label{eq:dp-in-Pi-dphi}
 \delta p&=& \frac{H}{a^3} \Pi - \frac{\dot{p}}{\dot{\rho}}\,6\, M_{\rm Pl}\,H^2\, \epsilon_1\, \left(\frac{\delta \phi}{\sqrt{2 \, \epsilon_1}}\right).
\end{eqnarray}
\end{subequations}
It is useful to identify the Bardeen potentials as
\begin{equation}
    \Psi = \psi - H \, \sigma = {\cal R} - H\, \Sigma\,, \quad \Phi = A + \dot{\sigma} = {\cal A} + \dot{\Sigma}.
\end{equation}
The momentum variable $\Pi$ can then be conveniently expressed as a Laplacian of $\Psi$ as
\begin{equation}\label{eq:Pi-Psi-1}
    \Pi = \frac{2 \,M_{\rm Pl}^2\, a^3 }{H} \frac{\nabla^2}{a^2} \Psi\,.
\end{equation}
Finally, the evolution of the comoving curvature perturbation and its conjugate momentum can be described by the following second-order differential equations:
\begin{subequations}\label{eq:eom-R-Pi}
\begin{eqnarray}
 \label{eq:eom-R}   {\cal R}'' + 2 \frac{z'}{z} {\cal R}' - \nabla^2 {\cal R} &=&0\, , \\
  \label{eq:eom-Pi}  \Pi'' - 2 \frac{z'}{z} \, \Pi' - \nabla^2 \, \Pi &=& 0\, , 
\end{eqnarray}
\end{subequations}
Introducing the Mukhanov–Sasaki variable $v \equiv z\, {\cal R}$ and its conjugate momentum $p \equiv \Pi/z$, the above equations can be equivalently rewritten as
 \begin{subequations}
 \label{eq:eom-v-p}
\begin{eqnarray}
 v'' - \left(\nabla^2+\frac{z''}{z} \right) v  &=&0, \\
    p'' - \left(\nabla^2+\frac{\theta''}{\theta} \right)p   &=& 0 , 
\end{eqnarray}
\end{subequations}
where $\theta \equiv 1/z$. It is important to emphasize that both $v$ and $p$ evolve homogeneously in space when the Laplacian terms are subdominant—specifically, in the regime where $\nabla^2 v \ll (z''/z) v$ and $\nabla^2 p \ll (\theta''/\theta) p$, respectively. 

\section{separate-universe Approach and Spatially-Homogeneous ${\cal R}$ and $\Pi$}\label{sec:background perturbations}
The conservation of the background energy density in an expanding universe is governed by the continuity equation:
\begin{equation}
    \dot{\rho} + 3 H \left( \rho + p\right) = 0.
    \label{eq:rho-bg}
\end{equation}
Now, to analyze the evolution of perturbations on top of this background, we introduce spatially homogeneous perturbations in the energy density, pressure, and time coordinate. Specifically, we consider~\cite{Wands:2000dp}
\begin{equation}
    \rho \rightarrow \rho + \delta \rho_{\rm h},\quad p \rightarrow p + \delta p_{\rm h},\quad {\rm d}t \rightarrow {\rm d}t(1 + A_{\rm h}). 
\end{equation}
where the subscript ``h" denotes the spatially homogeneous part of the perturbations. These substitutions effectively correspond to evaluating the continuity equation in a perturbed patch of the universe, i.e., in a separate ``local universe" approximation,
\begin{equation}
    \dot{\delta\rho}_{\rm h} + 3 H \left( \delta\rho_{\rm h} + \delta p_{\rm h}\right) + H \, \epsilon_1 \left( \delta \rho_{\rm h} +6 \, H^2 \, M_{\rm Pl}^2\,A_{\rm h} \right)= 0.
    \label{eq:rho-bg-perturbed}
\end{equation}
The first two terms in this equation take the same form as the unperturbed energy conservation equation, applied to the perturbed quantities. The third term arises due to the lapse perturbation $A$, which modifies the local time slicing. The combination $\delta \rho_{\rm h} +6 \,H^2 \,A_{\rm h}$ in the third term of~\ref{eq:rho-bg-perturbed} can be written as
\begin{equation}
    \delta \rho_{\rm h} +6 \, H^2 \, M_{\rm Pl}^2\, A_{\rm h} = \frac{H}{a^3} \Pi_{\rm h} + 6\, H\, M_{\rm Pl}^2\,\dot{\psi_{\rm h}}.
\end{equation}
It is easy to see that, using \ref{eq:Pi-Psi-1}, the quantity $\frac{H}{a^3} \Pi_{\rm h}$ in the above relation is related to spatial gradients through the Bardeen potential $\Psi$ as 
\begin{equation} \label{eq:Pi-Psi}
    \frac{H}{a^3} \Pi_{\rm h}= 2 \, M_{\rm Pl}^2\, \frac{\nabla^2}{a^2} \Psi\,.
\end{equation}
This term involves spatial gradients and is thus usually considered negligible on super-Hubble scales during standard slow-roll inflation. Consequently, the contribution of 
$\Pi$ to the local energy density perturbation is often ignored in the separate-universe approach. However, this approximation fails in more general scenarios. For example, during ultra-slow-roll inflation, the curvature perturbation ${\cal R}$ evolves non-trivially even on super-Hubble scales, and correspondingly, $\Pi$ does not vanish. In such cases, the gradient terms that are typically dropped in the separate-universe framework can become significant, leading to a breakdown of that approximation \cite{Jackson:2023obv}. Moreover, \ref{eq:Pi-Psi} suggests that the Bardeen potential $\Psi$ can exhibit non-local behavior over certain scales, allowing its Laplacian to become approximately homogeneous, and thus the corresponding $\Pi$ can appear spatially uniform~\cite{Cruces:2022dom}.

By substituting the expression for $\delta \rho$ from \ref{eq:drho-in-Pi-dphi}  into \ref{eq:rho-bg-perturbed} and using \ref{eq:Pi-in-sA} we get a simple relation as
\begin{equation}\label{eq:rho-bg-perturbed-Pi}
    \frac{H}{a^3}\dot{\Pi}_{\rm h} = 0.
\end{equation}
It is important to note that, in deriving the above relation, we have assumed $\Pi$ and $\delta \phi$ are spatially homogeneous. This approximation is valid during pure slow-roll and ultra-slow-roll inflation when the relevant length scales are much larger than the Hubble radius.

\subsection{Homogeneous solutions from linear perturbation theory}
In the last section, we have obtained equations governing comoving curvature perturbation and corresponding momentum and both are in general functions of time as well as space. From \ref{eq:eom-R} one can obtain
\begin{equation} \label{eq:Pi-dot-R}
\frac{\dot{\Pi}}{a^3} = 2\, M_{\rm Pl}^2\frac{\nabla^2 \left(\epsilon_1\,{\cal R}\right)} {a^2} .
\end{equation}
If the spatial gradient of $\epsilon_1\,{\cal R}$ is sufficiently small, and both ${\cal R}$ and $\Pi$ are spatially homogeneous, then the evolution equation of linear perturbations derived from the full Einstein equations reduces to the same form as~\ref{eq:rho-bg-perturbed-Pi}, which was obtained by perturbing the background energy conservation equation. This equivalence captures the essential idea behind the separate-universe approach: under the conditions of negligible spatial gradients and homogeneity of long-wavelength perturbations, the evolution of inhomogeneities in the universe can be effectively described by the evolution of locally homogeneous patches of space.

As mentioned earlier, the spatial homogeneity of ${\cal R}$ and $\Pi$ is essential for the separate-universe approach to remain valid. It is well established that in pure slow-roll and ultra-slow-roll inflation, this assumption holds on super-Hubble scales. The main goal of this work is to examine the breakdown of this assumption during transitions involving transient ultra-slow-roll phases. 


\subsection{Homogeneous solution of $\Pi$ and ${\cal R}$}
From \ref{eq:eom-Pi}, the homogeneous solution for $\Pi$ can be written as
\begin{equation}\label{eq:Pi-hom}
    \Pi_{\rm h} = \Pi_{i} + A \int_{N_i}^{N} \frac{z^2}{a\, H} \, {\rm d}{\bar N}\, ,
\end{equation} 
where $\Pi_{i}$ is the initial value of $\Pi$, and $A$ is a constant. Both $\Pi_i$ and $A$ are determined by the solution of the full (inhomogeneous) equation of motion \ref{eq:eom-Pi} evaluated at the initial time $N_i$. Furthermore, in Fourier space, the constant $A$ can be identified using \ref{eq:Pi-dot-R} as $A = -k^2\, \mathcal{R}_i$. This implies that the homogeneous solution of $\Pi$ inherently includes a contribution from the gradient of the comoving curvature perturbation $\mathcal{R}$.

In the case of power-law inflation, where the first slow-roll parameter $\epsilon_1$ is constant, the homogeneous solution for the conjugate momentum $\Pi$ evolves as
\begin{equation}
    \Pi_{\rm h} = \Pi_{i} + 2 \, A \frac{\epsilon_1}{H}\, \left(a - a_i\right).
\end{equation}
For $a \gg a_i$, we get
\begin{equation}
   H\,\frac{\Pi_{\rm h}}{a^3} = 2 \, A \frac{\epsilon_1}{a^2}\propto \frac{1}{a^2}.
\end{equation}
If the above contribution to $\delta \rho$ is absorbed into the background Hubble parameter, it can be interpreted as arising from an effective spatial curvature. This suggests that the gradient term contribution of $\mathcal{R}$ can be consistently incorporated into the background evolution and understood as a spatial curvature-like correction to the Hubble parameter. Such an interpretation is a key ingredient in the extended $\delta N$ formalism proposed in \cite{Artigas:2024ajh}.

For models in which $z^2$ decreases with time—such as in the case of ultra-slow-roll inflation—$\Pi_{\rm h}$ can be approximated as a constant value, as discussed in~\cite{Raveendran:2023auh}. In such scenarios, the quantity
\begin{equation}
   H\,\frac{\Pi_{\rm h}}{a^3} \propto \frac{1}{a^3}.
\end{equation}
This term behaves like the evolution of the energy density of the pressureless matter. This scaling behavior is analogous to the energy density evolution of pressureless matter.

The general solution for the comoving curvature perturbation can be expressed as
\begin{equation}
    {\cal R} = {\cal R}_{i} + \int_{N_i}^{N} \frac{\Pi_{\rm h}}{2\,M_{\rm Pl}^2\, a^3\, H\,\epsilon_1} \, {\rm d}{\bar N}.
\end{equation}
This expression indicates that, although ${\cal R}$ itself may be inhomogeneous, its evolution can be entirely reconstructed from the homogeneous component of $\Pi$. The corresponding homogeneous solution for ${\cal R}$ takes the form
\begin{equation}\label{eq: R-hom}
    {\cal R}_{\rm h} = {\cal R}_{i} + \Pi_{i}\, \int_{N_i}^{N} \frac{1}{2\,M_{\rm Pl}^2\, a^3\, H\,\epsilon_1} \, {\rm d}{\bar N}.
\end{equation}
Where ${\cal R}_{i}$ and $\Pi_{i}$ are supplied by solution of \ref{eq:eom-R-Pi}.
\section{Transient Ultra-Slow-Roll inflation}\label{sec:USR}
As mentioned earlier, we are interested in inflationary models that contain a transient ultra-slow-roll phase. Such models necessarily involve two transitions: first, from an initial slow-roll phase to ultra-slow-roll, and then from ultra-slow-roll back to a final slow-roll phase. Our aim is to study the evolution of ${\cal R}$ and $\Pi$ across these transitions and to examine the validity of using homogeneous solutions to trace their behavior on super-Hubble scales. As seen from \ref{eq:eom-v-p}, deviations from homogeneous evolution are controlled by the effective potentials $\nabla^2 + z''/z$ for ${\cal R}$ and $\nabla^2 + \theta''/\theta$ for $\Pi$.

Let us analyze the behavior of $z''/z$ and $\theta''/\theta$ near the transitions. During slow-roll phases, the quantity $z$ grows with time, as in both the initial and final slow-roll epochs of the models considered. By contrast, during the intermediate ultra-slow-roll phase, $z$ decreases with time. Thus, $z$ reaches a maximum at the first transition, while $\theta = 1/z$ reaches a maximum at the second transition. This implies that $z''/z$ becomes negative around the first transition and $\theta''/\theta$ becomes negative around the second. Consequently, $z''/z \, v$ (and $\theta''/\theta \, p$) inevitably cross $\nabla^2 \,v$ (and $\nabla^2 \,p$) during the respective transitions. Hence, the gradient terms become non-negligible in the evolution of ${\cal R}$ at the first transition and in the evolution of $\Pi$ at the second transition. Therefore, we expect the separate-universe approximation to break down at both transitions. In what follows, we show explicitly that this breakdown occurs in transient ultra-slow-roll models by numerically evolving the perturbations for a representative case. 

Typically, the models of interest are constructed either from a specific potential (see, e.g., Ref.\cite{Dalianis:2018frf}) or by parametrizing the slow-roll parameters directly \cite{Cai:2018dkf, Taoso:2021uvl, Franciolini:2023agm}. Since our focus is on understanding the impact of transitions on perturbation dynamics, it is convenient to adopt a framework where the sharpness and strength of transitions can be controlled explicitly. To this end, we consider a simple semi-analytical inflationary model in which the parameter $\eta \equiv \epsilon_1 - \epsilon_2/2$ is parametrized by a hyperbolic tangent function, following\cite{Franciolini:2023agm}:
	
	\begin{align}
	\eta(N)  & = \frac{1}{2}\left[-\epsilon_{2i}
	-\eta_{\rm II}+ \eta_{\rm II}\tanh\left(\frac{N-N_1}{\Delta N}\right)
	\right] \nonumber\\ 
 &+ \frac{1}{2}\left[
	\eta_{\rm II} + \eta_{\rm III} + (\eta_{\rm III}-\eta_{\rm II})\tanh\left(\frac{N-N_2}{\Delta N}\right)
	\right]\,,\label{eq:DynEta}
	\end{align}
where the parameter $\Delta N$ controls the width of the two transitions at $N_1$ and $N_2$. We choose following values for the parameters for our model:
    \begin{equation}
        N_1= 43.5\,,\, N_2 = 46\,, \epsilon_{2i}=\frac{1}{25}\,\,,  \eta_{\rm II}=3\,, \, \eta_{\rm III} = -0.7\,.
    \end{equation}
Addionally, in our model, the pivot scale $k_p = 0.05\, {\rm Mpc}^{-1}$ exits the Hubble radius at $N_p=10$ and the inflation ends at $N_e= 62$. This corresponds to a total number of e-folds between the Hubble exit of the pivot scale and the end of inflation $ N_\ast \equiv N_e - N_p = 52$. It should be noted that fixing the exact value of $N_\ast$ requires a detailed specification of the post-inflationary history, including the duration and equation of state during reheating, as well as the subsequent radiation-dominated era. These effects can shift $N_\ast$ by a few e-folds. Nevertheless, since the results presented in this work are insensitive to small variations in $N_\ast$, we adopt $N_\ast = 52$ as a representative value throughout our analysis. The constant $\epsilon_{2i} = 1/25$ ensures a spectral index of $n_s = 0.96$ on the CMB scales \cite{Planck:2018jri}. In this model, $k = 5.6 \times 10^{13} \, {\rm Mpc}^{-1}$ has the maximum power $\mathcal{P}_{\mathcal{R}} = 2\times 10^{-2}$. We numerically solve for the relevant background quantities and evolve the perturbations within the framework of the model described above. The perturbations are evolved in Fourier space, meaning that we solve the equations mode by mode for each wavevector $k$. In our analysis, we focus on the specific mode $k = 10^{12} \, {\rm Mpc}^{-1}$, which exits the Hubble radius just before the transition from the initial slow-roll phase to the ultra slow-roll phase. The results of this numerical evolution and their implications are discussed in the following paragraphs.

\begin{figure}[!]
\centering
\includegraphics[width=0.95\linewidth]{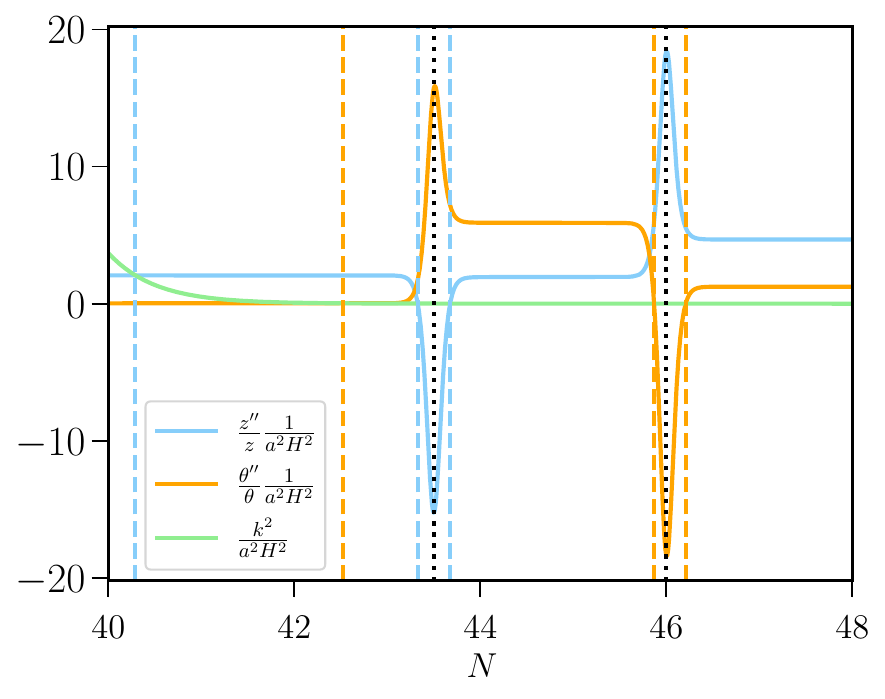}
\caption{Evolutions of $z''/(z \,a^2 \,H^2) $, $\theta''/(\theta \,a^2 \,H^2)$ and $k^2/(a^2 \,H^2)$ as functions of the number of e-folds $N$. The black dotted vertical lines indicate the two transition points in the background evolution. The blue dashed vertical lines mark the moments when $k^2 = z''/z$, while the orange vertical lines indicate when $k^2 = \theta''/\theta$. The figure shows that the evolution of ${\cal R}$ crosses the $z''/z$  and becomes affected by the $k^2$ term twice during the first transition. Similarly, the evolution of $\Pi$ crosses $\theta''/\theta$ and becomes influenced by the $k^2$ term twice during the second transition. These crossings signal the breakdown of the separate-universe approximation during the respective transition phases.}\label{fig:zppz-usr}
\end{figure}
\ref{fig:zppz-usr} shows the evolution of $z''/z$, $\theta''/\theta$, and $k^2$. It is evident from the figure that $k^2$ crosses $z''/z$ twice during the first transition, in agreement with the analytical expectations.. This implies that the gradient term $k^2$ cannot be neglected in the evolution of the comoving curvature perturbation ${\cal R}$ during this phase. Similarly, $k^2$ crosses $\theta''/\theta$ twice during the second transition, indicating that its contribution cannot be ignored in the evolution of the conjugate momentum $\Pi$ during that phase. As mentioned earlier, for the separate-universe approach to remain valid, both ${\cal R}$ and $\Pi$ must evolve homogeneously. Therefore, we conclude that the separate-universe approximation breaks down during both the first and second transitions.

\begin{figure}[!]
\centering
\includegraphics[width=0.95\linewidth]{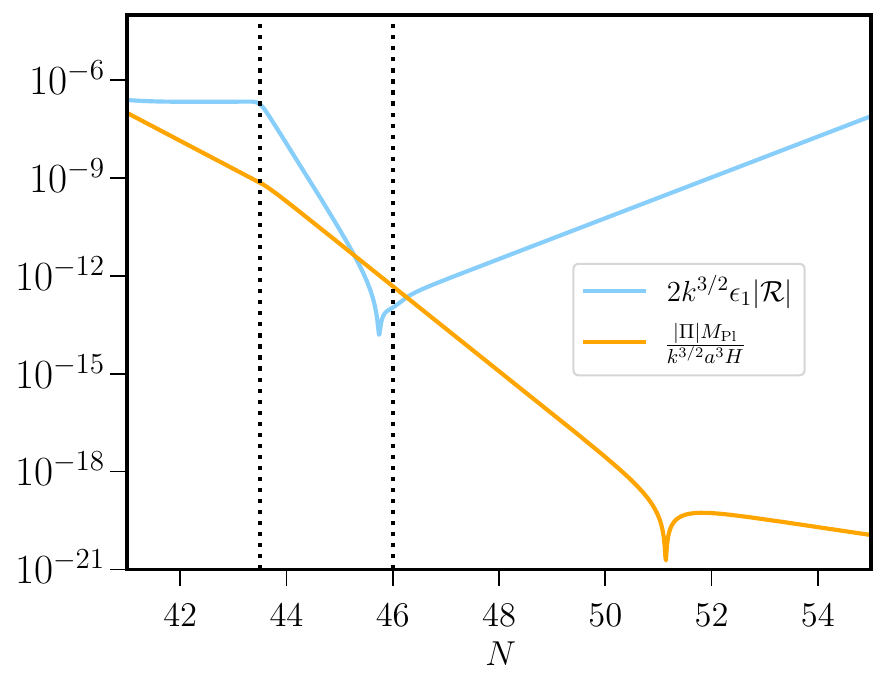}
\caption{Evolutions of $2 \, \epsilon_1 k^{3/2}\vert{\cal R}\vert$ and $\vert\Pi\vert\, M_{\rm Pl}/(k^{3/2}a^3\, H\, )$ in Fourier space. The figure clearly demonstrates that the momentum term becomes dominant and cannot be neglected during the ultra-slow-roll phase in the expression for $\delta \rho$ given in~\ref{eq:drho-in-Pi-dphi}.}\label{fig:rho-terms}
\end{figure}
\ref{fig:rho-terms} illustrates the evolution of the contributions to the energy density perturbation $\delta \rho$ from the comoving curvature perturbation in the spatially flat gauge, given by $\epsilon_1\, k^{3/2}\,\vert{\cal R}\vert$, and from the momentum term, $\vert\Pi\vert \, M_{\rm Pl}/(k^{3/2}\,a^3\,H)$. During the slow-roll phase, the momentum contribution is clearly subdominant compared to that of ${\cal R}$, justifying its common neglect in the separate-universe approximation. However, the figure shows that in the ultra-slow-roll regime, the momentum term becomes dominant and cannot be ignored, as it constitutes the leading contribution to $\delta \rho$. This dominance arises because $\Pi$ remains approximately constant during this phase, leading to a scaling behavior of $\Pi/(a^3\,H) \propto 1/a^3$.
\begin{figure}[!]
\centering
\includegraphics[width=0.95\linewidth]{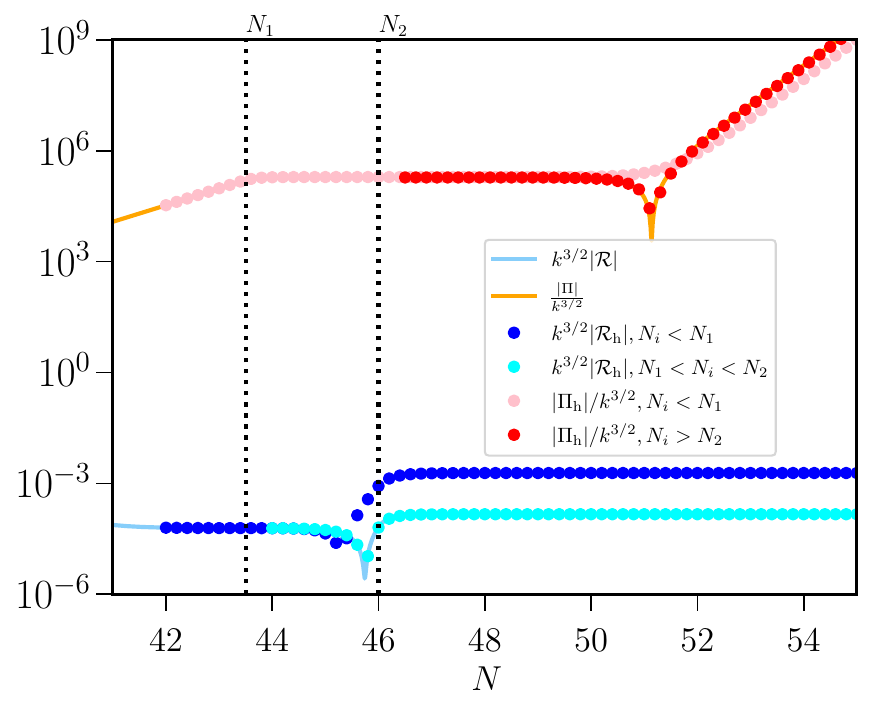}
\caption{Evolutions of $\vert{\cal R}\vert$ and $\vert\Pi\vert$ are shown as solid lines, obtained by numerically solving \ref{eq:eom-R-Pi}. These are plotted alongside the corresponding homogeneous solutions, initialized at different times, represented by data points. The light-blue solid line denotes the exact evolution of ${\cal R}$, while the blue data points represent the homogeneous solution ${\cal R}_{\rm h}$ when initial conditions are set before the first transition. The cyan data points show ${\cal R}_{\rm h}$ when initialized after the first transition. Similarly, the orange solid line indicates the exact evolution of $\Pi$. The pink data points correspond to the homogeneous solution $\Pi_{\rm h}$ initialized before the first transition, and the red data points correspond to $\Pi_{\rm h}$ when initialized after the second transition. }\label{fig:R-Pi}
\end{figure}

\ref{fig:R-Pi} shows the evolution of $\vert{\cal R}\vert$ and $\vert\Pi\vert$ obtained by numerically solving~\ref{eq:eom-R-Pi} (solid lines), along with the corresponding homogeneous solutions initialized at different times (shown as solid circular data points). The plot illustrates that the homogeneous solutions for ${\cal R}$ and $\Pi$ do not fully capture their evolution throughout the entire inflationary period when initial conditions are set during the first slow-roll phase. For ${\cal R}$, the homogeneous solution accurately describes its behavior only during the first slow-roll phase, breaks down during the transition, and becomes valid again after the transition if the initial conditions are redefined. In contrast, $\Pi$ follows the homogeneous solution during both the first slow-roll and the ultra-slow-roll phases, but deviates during the second transition. After this, it again matches the exact solution when reinitialized with updated initial conditions.

\section{Conclusion}\label{sec:summary}

Recently, it has been shown that the separate-universe approach transiently breaks down during the transition from slow-roll to ultra-slow-roll inflation~\cite{Leach:2001zf,Jackson:2023obv}. To extend its applicability through such transitions, a modified framework was proposed in~\cite{Artigas:2024ajh}. In this work, we revisit the breakdown of the separate-universe approach in transient inflationary scenarios and examine the proposed modifications by analyzing the evolution of the comoving curvature perturbation ${\cal R}$ and its conjugate momentum $\Pi$.

Our analysis leads to two main findings. First, the separate-universe approximation breaks down during both the slow-roll to ultra-slow-roll transition and the subsequent return to slow-roll. This failure is due to the non-negligible contribution of spatial gradient terms, which temporarily affect the evolution of ${\cal R}$ and $\Pi$. As shown in~\ref{eq:drho-in-Pi-dphi}, the perturbation in energy density receives contributions from both ${\cal R}$ (in the spatially flat gauge) and $\Pi$. For the separate-universe approximation to hold, both quantities must evolve homogeneously. However, our results show that ${\cal R}$ becomes inhomogeneous during the transition from slow-roll to ultra slow-roll, while $\Pi$ becomes inhomogeneous during the transition back from ultra slow-roll to slow-roll—thereby violating a key assumption of the separate-universe framework.

It is also important to clarify that the non-negligible contribution of spatial gradient terms to the super-Hubble evolution of ${\cal R}$ and $\Pi$ during the transitions is a generic feature of transient ultra-slow-roll inflationary models. However, the magnitude of this contribution, and consequently the extent to which the separate-universe approximation breaks down, depends on the model parameters that govern the strength of the transition. In this work, we have adopted a particular parametrization of $\eta$ to construct a simple transient ultra-slow-roll model and numerically evolve the perturbations to illustrate our claims. Nevertheless, one could equally well consider alternative realizations of transient ultra-slow-roll inflation tailored to different physical requirements. The central point of our analysis is that the separate-universe approximation generically breaks down at both transitions due to the presence of an intermediate ultra-slow-roll phase between two slow-roll phases, independent of the model-building details.

Second, we find that the evolution of the conjugate momentum $\Pi$ remains smooth and largely unaffected by spatial gradient terms after Hubble exit – not only during the initial slow-roll phase, but also during the transition from slow-roll to ultra-slow-roll, and throughout the ultra-slow-roll regime. Since this contribution is independent of gradients, it can be consistently absorbed into the background energy density, allowing one to define a modified Hubble parameter. Remarkably, this contribution scales as $1/a^2$ and, when the first slow roll parameter $\epsilon_1$ is small and constant, it can be interpreted as an effective spatial curvature term. This interpretation forms the basis of the extended $\delta N$ formalism proposed in~\cite{Artigas:2024ajh}.

However, we also find that this curvature-like interpretation holds strictly only under the assumption of a constant $\epsilon_1$, which is not generally satisfied in realistic inflationary scenarios. When this condition is relaxed, the contribution from $\Pi$ can still be absorbed into the background to define a modified local Hubble rate, but its identification with a spatial curvature term is no longer valid in the standard geometric sense.

We also demonstrate that the contribution of the momentum term to the energy-density perturbation is negligible during the slow-roll phases and is therefore typically ignored in the standard separate-universe approach. However, it becomes significant during the ultra-slow-roll phase, consistent with the findings of~\cite{Cruces:2022dom}. This suggests that it is important to track the evolution of the conjugate momentum in general inflationary models. In other words, adopting a Hamiltonian formalism—based on the comoving curvature perturbation and its conjugate momentum—can provide deeper insight into models that include phases deviating from the slow-roll approximation. Such a formulation may also lead to a version of the separate-universe approach within a Hamiltonian framework. Exploring this direction is part of our future work.

\acknowledgements
The author wish to thank Vincent Vennin and Cristiano Germani for interesting discussions. RNR is supported by the National Post-Doctoral Fellowship of Anusandhan National Research Foundation(ANRF), Science and Engineering Research Board (SERB), Department of Science and Technology (DST), Government of India (GOI)(PDF/2023/001226).

	
	

	

	\bibliographystyle{apsrev4-1}
	\bibliography{mybibliography-cosmology,extended-deltaN}

\end{document}